# You are now an Influencer! Measuring CEO Reputation in Social Media

**Julian Marx**
Professional Communication in Electronic Media/Social Media
University of Duisburg-Essen
Duisburg, Germany
Email: julian.marx@uni-due.de

**Milad Mirbabaie**
Professional Communication in Electronic Media/Social Media
University of Duisburg-Essen
Duisburg, Germany
Email: milad.mirbabaie@uni-due.de

**Stefan Stieglitz**
Professional Communication in Electronic Media/Social Media
University of Duisburg-Essen
Duisburg, Germany
Email: stefan.stieglitz@uni-due.de

## Abstract

We know that reputation in organisational contexts can be understood as a valuable asset that requires diligent management. It directly affects how a firm is publicly perceived, and indirectly, how a firm will perform economically. The establishment of social media as ubiquitous tools of communication have changed how corporations manage their reputation. Particularly CEOs face novel responsibilities, as they deal with their personal image, which at the same time affects the reputation of their firm. Whereas CEO and corporate reputation have been researched isolated from each other, little is known about how a CEO's social media reputation management affects corporate reputation. This research in progress paper aims to emphasise this research gap with a literature review on the current status of reputation management and measurement by means of social media. We further propose a research design that combines sentiment analysis, frequency detection, and content analysis and discuss further research prospects.

# 1　Introduction

Reputation in an organisational context is noted as an intangible, yet valuable asset that requires management as it directly impacts the stakeholders' perception of a firm (Puncheva 2008). A favourable reputation is a crucial objective for a company. From a business-to-consumer perspective, a positive reputation can offer advantages such as enabling firms to charge higher prices (Klein and Leffler 1981; Roberts and Milgrom 1986), attract more professional applicants (Stigler 1962) and incentivise private investors (Beatty and Ritter 1986). Eventually, a combination of those factors may result in increased financial performance (Deephouse 2000). Scholars have defined corporate reputation as a "perceptual representation of a firm's past actions and future prospects that describes the firm's overall appeal to all of its key constituents when compared with other leading competitors" (Fombrun 1996, p. 72).

The advent of social media has compelled the market to rethink the way in which it organizes business communication (Trainor 2012). As a consequence, reputation management has partly moved from the traditional news media settings towards communication through social media, where customers and patrons can be reached conveniently and cost efficiently (Stieglitz et al. 2014). From a management perspective, firms now face the challenge to oversee and guide these opinions to build, maintain, or protect a favourable reputation. It is therefore imperative that executive managers learn how to steer corporate reputation using social media strategies (Floreddu et al., 2014). This applies not only to their employer's reputation, but also to their own image. CEOs are often perceived as the professional face of the company and an expert in their field (Kietzmann et al. 2011). Scholars have acknowledged that a CEO's reputation affects not only their own career opportunities, but also the reputation of their firm (Hayward et al. 2004; Pamuksuz and Mourad 2016). However, little is known about how a CEO's reputation management in social media affects corporate reputation. Previous studies have concerned either CEO reputation in traditional news media settings (Deephouse 2000), or corporate reputation in social media (Benthaus 2014). Literature about CEO reputation management in social media is immensely underrepresented. Hence our research in progress pursues following research questions:

***RQ1***: *To what extent do dimensions of corporate reputation apply to CEO reputation in social media?*
***RQ2***: *How does CEO reputation in social media deviate from corporate reputation?*

In order to approximate a complete study on the above issue, we have conducted a literature review to map out the state of the art. This research in progress paper aims to put our current status up for discussion, and eventually form the basis of a comprehensive data-driven case study.

The paper is structured as follows: In section 2, we present our literature review on reputation management in social media, with particular consideration of CEOs. Subsequently, in section 3, we propose our research design, which consists of mixed methods to determine reputation indicators using social media data. In section 4, we reflect upon our current progress and share ideas for further research.

# 2　Background
## 2.1　Corporate Reputation Management in Social Media

Reputation covers polarised public opinions about a person or organisation. From a strategic point of view, it can be a valuable non-material asset (Hall 1992) and may shape an enterprise's approach on how to orient further actions (Wernerfelt 1984). Being positively perceived is difficult to imitate by competitors, therefore a strong reputation establishes and maintains competitive advantages (Madhani 2010). Potential customers view it as a crucial factor for the selection of their supplier (Walsh et al. 2009), and are willing to pay more for products and commodities (Dijkmans et al. 2015). Moreover, a positive corporate reputation creates market entry barriers for competitors, nurtures customer loyalty and retention (Nguyen and Leblanc 2001). This enables a firm to attract a wider customer base (Fombrun et al. 2000), which consequently translates into higher earnings (Smith et al. 2010). Stakeholders show increased willingness to purchase company shares, since a good reputation enables the firm to attract higher quality employees and to gain better returns (Chun 2005; Vergin and Qoronfleh 1998). In times of crisis, reputational capital holds the capacity to protect a company (Shamma 2012). According to Fombrun et al. (2000), reputation is a dyadic concept, as it consists of an emotional (affective) and a rational (cognitive) component. Therefore, as shown in figure 1, the authors subdivided corporate reputation into six dimensions, which mirror the spectrum of corporate communication on a content-level.

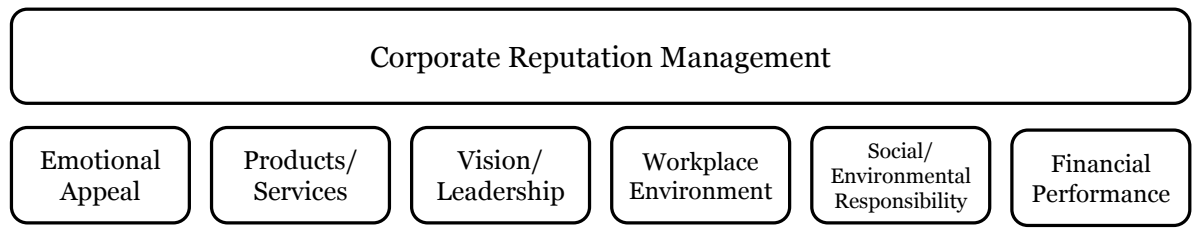

*Figure 1: Dimensions of Corporate Reputation (Fombrun et al. 2000)*

These dimensions provide a framework for approaching the key components of corporate reputation. Conveying *Emotional Appeal* in a communication strategy may result in positive feelings and respect for the company and eventually increase trust. Presenting *Products and Services* is rather oriented toward marketing and holds the capacity to lead consumers to perceiving an organisation as innovative, expecting high product quality, or to express identification with those products and services. *Vision and Leadership* represents the corporate mission as well as a goal-oriented execution of a company's activities. The impression of a firm to be a popular workplace is part of the Workplace Environment dimension. Addressing *Social/Environmental Responsibility* reflects a firm's commitment to good causes and responsibility towards environment and society. *Financial Performance* indicates an organization to be profitable, and to be capable to outperform competitors. Therefore, it is expected to grow and be of value to shareholders. Scholars have transferred those dimensions into social media settings as they have proven to be eligible segments to categorise social media content (Benthaus 2014). Kaplan and Haenlein (2010) noted that "social media allow firms to engage in timely and direct end-consumer contact at relatively low cost and higher levels of efficiency than can be achieved with more traditional communication tools" (p.67). Facebook, Instagram or Twitter are among the most popular social media platforms for those interactions (Berthon 2012; Funk 2011). Karjaluoto et al. (2016) analysed the relationship between a company's social media reputation and firm performance. Results indicate that companies with an active social media presence tend to have better reputations than companies forgoing social media. As a matter of fact, organisations focus on promoting themselves positively on such channels to support their reputation (Aral et al. 2013). Entering social media as a novel realm of corporate communication comes with several challenges, e.g. the bidirectional communication with the public and low control over what is part of public conversations (Karjaluolto, Mäkinen and Järvinen 2015). A study of Stieglitz et al. (2018a) investigated the reputation management in social media of VW during the rise of the "Dieselgate" scandal. Their results revealed that VW kept silent instead of actively managing their reputation. This emphasises missing confidence to be able to control public opinions and the need for empowering firm's communication strategies. Utilising social media for reputation management results in higher volatility of reputation, as opinions, ideas, or contradictory viewpoints circulate in online social networks with amplified velocity (Dijkmans et al. 2015). Hence, corporate reputation becomes a much more fragile factor than in the pre-social media era. The unexpected virality of user-generated content may either be a blessing or a curse for a company's reputation (Colleoni et al. 2011).

## 2.2　CEO Reputation Management in Social Media

With increasing popularity of social media platforms, companies not only appear with their corporate branding, but also personified in brand pages of their managers, in particular CEOs (Aral et al. 2013). Same as companies, CEOs are entities which seek to convey a positive image of themselves (Mohamed et al. 1999). Through individual social media profiles, CEOs do not only transmit expertise but can also resonate sympathy and become approachable to their stakeholders (Pamuksuz and Mourad 2016). Personal reputation building enables an emotional connection with others, a mechanism which works for CEOs to compete in the internal hierarchy and the external job market (Arruda and Dixson 2007; Chen et al. 2015). CEO reputation differs from corporate reputation as they establish themselves independently. A personal image already exists solely through social interaction and a career prior to the employment in the current profession (Rangarajan et al. 2017). On the one hand, positive personal reputation can be empowering for their managerial authority, and thus, increase their career prospects (Hayward et al. 2004). But on the other hand, negative reputation might reduce managerial power and harms their image (Wiesenfeld et al. 2008). According to Mohamed et al. (1999) there are distinctive tactics how managers balance their reputation, also referred to as *impression management*. Assertive and defensive tactics deal with the active management of positive or negative perceptions. Disclosing private information that mirrors personal viewpoints or the social connection to others are represented as direct and indirect tactics. In general, CEOs utilise these tactics to shape their own image and minimise the risk of a harmful reputation. As a result, executives such as CEOs experience great responsibility as their personal reputation might not only impact their career, but also the public impression of their firm.

Reputation management of CEOs in social media differs from a traditional media setting. CEOs may pursue an individual publishing strategy, e.g. to regulate the frequency of public appearances rather than depending on editorship. Social media allows them constantly communicate and network on a more customised level (Pamuksuz and Mourad 2016). The impact of social media logics applied to CEOs' reputation poses a number of crucial questions to be examined. Kietzmann et al. (2011) presume a positive CEO reputation to be highly associated with the standing of an executive in a social network. Apart from that, little is known about what metrics define a CEOs reputation in social media. Previous studies have concerned CEO reputation in traditional media settings. Deephouse (2000), for instance, examined the consequences of reputation from a resource-based perspective, such as financial performance and executive compensation, using the volume of coverage in news media as a reputation score. Francis et al. (2008), too, examined CEO reputation in traditional print media and revealed an impact on the earnings quality of a company. However, current literature does not provide sufficient dimensions to characterise the role of CEOs in social media settings, as well as consideration for their impact on corporate reputation.

## 3   Proposed Methods for Researching Reputation in Social Media

In order to approximate an eligible research design to measure CEO reputation and its impact on corporate reputation, we propose a single case study design. The comparative element lies in the superimposition of (1) corporate reputation and (2) CEO reputation over a time period of 16 weeks to oversee short-term volatility. On a conceptual level, we turn to Yin (2003) for an explorative design and to Hays (2004) for validation, e.g. through triangulation. The case should involve a large corporation (e.g. listed as Fortune 500) with corporate and CEO social media branding (>100.000 platform-specific followers). An ideal timing would presuppose a corporate crisis setting as reputation is often socially re-evaluated during times of corporate crisis (Deephouse and Suchman 2008). Our study employs a mixed methods approach, including quantitative sentiment analysis and qualitative content analysis, as shown in figure 2.

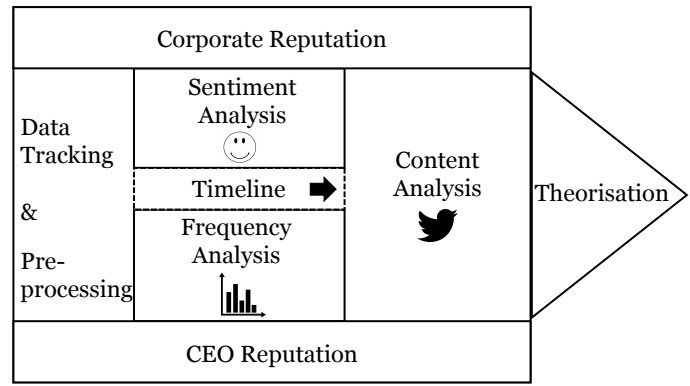

*Figure 2: Framework for reputation measurement using social media data*

### 3.1   Sentiment Analysis

To determine the conversational tone about CEOs and their apposite corporation, we suggest conducting sentiment analyses as an initial point of our study. To this end, we require a dataset consisting of a representative set of social media postings. Twitter qualifies as a suitable source of data due to the fact that it frequently hosts bidirectional communication between public figures or organisations and private individuals (Stieglitz et al. 2018b). Moreover, thanks to an open search API, Twitter data is publicly available for research endeavours and other purposes. Using a self-developed Java crawler in combination with an open source library (e.g. Twitter4J) allows it to obtain large tweet samples. Collected data may include original postings as well as forwarding messages that include relevant keywords pointing to the subjects of research, e.g. the name of the CEO/company or synonymous brands or expressions. Extracted data subsequently need further (pre-processing) to be suited for sentiment analyses. An automated analysis with SentiStrength[1] yields values for positive sentiment (1 to 5) and negative sentiment (-1 to -5). This lexicon-based approach allows to show strong indicators for the overall sentiment of a tweet. As an additional metric, the frequency of mentions should be taken into account. This includes the mere occurrence of keywords as well as Twitter mentions (@username). Using this frequency analysis as an indicator for reputation transfers the approach of Deephouse (2000) onto a social media setting. Once sentiment and frequency are determined, a visualisation in form of two parallel timelines promises valuable findings. If overlapping sentiments follow a similar graph, one could draw conclusions about the correlation of corporate and CEO reputation.

---

[1] http://sentistrength.wlv.ac.uk/

### 3.2 Content Analysis

To achieve a more valid index of reputation and to understand possible fluctuations in the sentiment and frequency graphs, we propose a content analysis of a tweet sample. The tone of voice in public media as measured by means of a sentiment analysis is crucial to assess reputation. However, there are additional dimensions to it such as leadership, strategy, culture, and innovation (Cravens et al. 2003). Following the guidelines of Mayring (2000, 2014), categories for the qualitative assessment of opinions about the corresponding firms may be derived from Fombrun et al. (2000) and Cravens et al. (2003). Since our study approaches the measurement of reputation from a social media data perspective, we target to grasp those dimensions through conducting a content analysis of tweets. If the same categories are applied for classifying the mentions of CEOs, the analysis may reveal whether the dimensions of corporate reputation suit CEO reputation or if additional categories are needed. On Twitter, especially the retweet count indicates the reach and impact of a message (Pal and Counts 2011). Therefore, a ranking of most retweeted postings qualifies as a selection procedure for relevant tweets to build a subsample.

## 4   Conclusion and Further Research

This research in progress paper aims to lay the foundation for a study on CEO reputation management in social media. At the current stage of our project, we propose a mixed method design to approach our research questions. First, a sentiment analysis should be employed to determine the tone of public opinions about a corporation and its CEO, respectively. This analytical step should incorporate the frequency of mentions following the groundwork of Deephouse (2000). In a subsequent step, a content analysis of representative tweets test existing dimensions of corporate reputation (Fombrun et al. 2000) with regards to their applicability for CEO reputation. From analysing sentiments within our case study, we expect findings that mirror CEO reputation to be unattached to the reputation of the respective corporate brand. The textual analysis, however, would allow us to determine additional dimensions of CEO reputation. We expect those dimensions to be based upon the individual characteristics of a CEO as opposed to a corporate brand, e.g. personality, authority, or authenticity.

The proposed study design comes with limitations as a method mix may lead to more valid results, but does not guarantee the appropriateness of this mix for measuring reputation. However, using sentiment analysis allows to process larger data samples than, for instance, surveying a sample customers to assess reputation. Our intended research design suggests Twitter as a suitable platform to retrieve data from. However, going beyond a single data source might enhance the significance of results and enables researchers to gain a broader view on reputation management strategies and a holistic picture of circulating opinions in social media.

Our suggestions for further research are to determine a suitable set of company and CEO pairs who are actively building their reputations in social media. Current best practices of high executives in social media, we argue, show lots of unused potential. Addressing CEO-specific dimensions in social media reputation management may endow CEOs with the standing of opinion leaders. Both their corporation and personal career may reap benefits from such social standing.